\documentclass[preprint]{elsarticle}    

\usepackage{graphicx}          
\usepackage{amsmath}
\usepackage{amssymb}
\usepackage{graphicx}
\usepackage{graphicx}
\usepackage{dcolumn}
\usepackage{bm}

\newcommand{\be}{\begin{eqnarray}}
\newcommand{\ee}{\end{eqnarray}}

 \newcommand{\gsim}{\mathrel{\hbox{\rlap{\lower.55ex \hbox {$\sim$}}
                   \kern-.3em \raise.4ex \hbox{$>$}}}}
\newcommand{\lsim}{\mathrel{\hbox{\rlap{\lower.55ex \hbox {$\sim$}}
                   \kern-.3em \raise.4ex \hbox{$<$}}}}

\newcommand{\vz}{\bar v_z}
\newcommand{\tp}{\tau_\perp}
\newcommand{\gm}{\bar \gamma}
\newcommand{\ka}{\kappa}

\begin{document}

\begin{frontmatter}

\title{On Analytic Solutions of (1+3)D Relativistic Ideal Hydrodynamic Equations}

\author[sl]{Shu Lin \fnref{fn1}}
\author[jl]{Jinfeng Liao \fnref{fn2}}

\address[sl]{Department of Physics \& Astronomy, SUNY Stony Brook, NY11794, USA.}

\address[jl]{Nuclear Science Division, Lawrence Berkeley National Laboratory, \\
MS70R0319, 1 Cyclotron Road, Berkeley, CA 94720, USA.}

\fntext[fn1]{slin@grad.physics.sunysb.edu}
\fntext[fn2]{jliao@lbl.gov}

\begin{abstract}
In this paper, we find various analytic (1+3)D solutions to
relativistic ideal hydrodynamic equations based on embedding of
known low-dimensional scaling solutions. We first study a class of
flows with 2D Hubble Embedding, for which a single ordinary
differential equation for the remaining velocity field can be
derived. Using this equation, all solutions with transverse 2D
Hubble embedding and power law ansatz for the remaining
longitudinal velocity field will be found. Going beyond the power
law ansatz, we further find a few solutions with transverse 2D
Hubble embedding and nontrivial longitudinal velocity field.
Finally we investigate general scaling flows with each component
of
 the velocity fields scaling independently, for which we also find all
 possible solutions.
\end{abstract}



\end{frontmatter}

\newpage

\setcounter{page}1

\section{Introduction}

Relativistic hydrodynamics \cite{Landau_book,Weinberg_book} is
widely used in physics. Its particular application to high energy
collision dynamics can be dated back to Laudau and Khalatnikov
\cite{Landau_Khalatnikov} in 1950's. For a brief history and an
introduction to this aspect, see e.g. \cite{ES_book}. The
relativistic ideal hydrodynamics (RIHD) has played a significant
role in the past decade for the development of theory and
phenomenology of the quark-gluon plasma (QGP)
\cite{Shuryak:1978ij} --- a deconfined high temperature phase of
QCD matter. The QGP can be experimentally created and studied by
the Relativistic Heavy Ion Collisions (RHIC).  The space-time
evolution at RHIC is characterized by fast longitudinal expansion
(longitudinal flow) and strong transverse expansion (radial and
elliptic flow). In particular, the large elliptic flow $v_2$ in
non-central collisions \cite{Ollitrault:1992bk} can be well
described by RIHD \cite{Teaney:2000cw} at low-to-intermediate
transverse momenta for almost all particle species and for various
centralities, beam energies and colliding nuclei, implying very
low viscosity of the underlying matter, see reviews in e.g.
\cite{Voloshin:2008dg,Heinz:2009xj}. It is now widely believed
that the QGP in $1-2T_c$ region (as covered by current RHIC
energy) is a ``nearly perfect fluid''
\cite{Schaefer:2009dj}\cite{Liao:2009gb} (see however caveats as
discussed in e.g. \cite{Liao:2009ni}\cite{Koch:2009wk}). The
success of RIHD description for RHIC space-time evolution has
become the basis for the paradigm shift that the QGP in $1-2T_c$
region is actually a strongly-coupled liquid
\cite{Shuryak_review,sQGP_review}. Such strong coupling and good
fluidity close to $T_c$ may have its origin in the magnetic
component of QCD plasma \cite{Liao:2006ry,Chernodub:2006gu} which
is ultimately connected with the mechanism of QCD deconfinement
transition.

 The success of RIHD has also
inspired a lot of interest in the formal study of RIHD. Analytic
solutions from RIHD have been very useful for RHIC phenomenology,
for example the well-known Hwa-Bjorken boost invariant flow
\cite{Hwa,Bjorken}. There have been many efforts and progresses in
finding analytic solutions, with an emphasis on solutions with
potential importance for describing RHIC collisions. Study aiming
at (1+1)D solutions that give an alternative description of the
longitudinal expansion and a more realistic (non-boost-invariant)
multiplicity distribution over rapidity can be found in e.g.
\cite{Nagy:2007xn,Bialas:2007iu,Pratt:2008jj,Wong:2008ex}. General
Solutions of the (1+1)D RIHD with arbitrary initial conditions in
the forward light-cone for specific equation of states are also
found in \cite{Nagy:2007xn}.  Beyond (1+1)D, ellipsoidally
expanding flows have been studied in
\cite{Csorgo:2002bi}\cite{Sinyukov:2004am}, and generalization of
the spherically symmetric Hubble flows to accelerating, non-Hubble
flows in arbitrary (1+d)D has been done in \cite{Nagy:2007xn}.
There have also been study of adding radial flow to a longitudinal
Bjorken profile in e.g. \cite{Biro:1999eh}, and more recently
attempts to deal with elliptic flow analytically
\cite{Peschanski:2009tg}. A new method for solving relativistic
ideal hydrodynamics in (1+3)D is developed in \cite{Liao:2009zg},
where by explicitly embedding in the longitudinal and transverse
radial flows, the authors reduce the RIHD equations to a single
equation for the transverse velocity field only and find
analytically all possible solutions with their transverse velocity
fields having power dependence on proper time and transverse
radius.


In this paper, we aim to find analytic solutions to RIHD in
(1+3)D, by taking the approach of Liao and Koch in
\cite{Liao:2009zg}. Basically one uses lower-dimensional scaling
solutions (Hubble flow in 1D and 2D) to build solutions in
higher-dimension. We will first study a class of flows with 2D
Hubble Embedding, for which a single ordinary differential
equation for the remaining velocity field will be derived in
Section-3. Using this equation, all solutions with transverse 2D
Hubble embedding and power law ansatz
 for the remaining velocity field will be found in Section-4.
More interesting solutions with transverse 2D Hubble embedding and
nontrivial longitudinal flow beyond power law ansatz will be
investigated and found in Section-5. In Section-6 we will study
general scaling flows with each component of the velocity fields
scaling independently, for which we also find all possible
solutions. Finally we summarize and discuss the results in
Section-7.

\section{Generalities}

We start with the relativistic hydrodynamics equations:
\begin{equation}\label{eqn_hydro}
T^{m n}\,  _{; \, n}=0
\end{equation}
For ideal hydrodynamics, the energy-momentum tensor is given by
\begin{equation} \label{eqn_tmunu}
T^{m n} = (\epsilon + p)\, u^m u^n - p \, g^{m n} = w \, u^m u^n -
p \, g^{m n}
\end{equation}
with $\epsilon,p$ and $w=\epsilon+p$ being the energy density,
pressure and the enthalpy density defined in the flowing matter's
local rest frame (L.R.F) which by definition are Lorentz scalars.
The flow field $u^{m}(x)$ is constrained by $u^m\cdot u_m =1$. In
the usual $(t,\vec x)$ coordinates one can express $u^{m}(x)$ as
$\gamma (1,\vec v)$ with $\gamma=1/\sqrt{1-\vec v^2}$ and $\vec
v=d\vec x/dt$.

To close the set of equations for five independent fields
$\epsilon,p,\vec v$, we also need Equation of State (E.o.S)
\begin{equation} \label{eqn_eos}
 p=\nu (\epsilon+p) = \nu w
\end{equation}
The above means $\epsilon=\frac{1-\nu}{\nu} p$ implying a speed of
sound $c_s=\sqrt{\partial p \over
\partial \epsilon}=\sqrt{\frac{\nu}{1-\nu}}$, and, in order to assure $c_s\le 1$,
we require $0<\nu\le 1/2$.

\subsection{Curved coordinates for 2D Hubble Embedding}
For a 2D Hubble Embedding study, we will use a coordinate system
of $(\tau_\perp,\eta_\perp,\phi_\perp,z)$: i.e. the transverse
proper time, the spatial transverse rapidity, the transverse
angle, and the longitudinal $z$. They are related to the usual
$(t,x,y,z)$ in the following way (with
$r_\perp\equiv\sqrt{x^2+y^2}$) :
\begin{eqnarray}
&& \tau_\perp=\sqrt{t^2-r_\perp^2} \,\, , \,\,
\eta_\perp=\frac{1}{2}\mathbf{ln}\frac{t+r_\perp}{t-r_\perp}\, ,
\nonumber \\
&& \phi_\perp=\frac{1}{2i}\mathbf{ln}\frac{x+y\cdot i}{x-y\cdot i}
\,\, , \,\, z=z
\end{eqnarray}
and inversely
\begin{eqnarray}
&& t=\tau_\perp \cosh \eta_\perp \,\, , \,\, z=z \,\, , \,
\nonumber \\
&& x=\tau_\perp \sinh \eta_\perp \cos\phi_\perp \,\,  , \,\,
y=\tau_\perp \sinh \eta_\perp \sin\phi_\perp
\end{eqnarray}

The velocity field $u^{m}$ in these coordinates is related to
$u^{\mu}=\gamma(1,v_x,v_y,v_z)$ in flat coordinates $(t,x,y,z)$
via
\begin{eqnarray} \label{eqn_curved_v}
&& u^{\tau_\perp} = \gamma [ \cosh \eta_\perp - (v_x\cos
\phi_\perp + v_y \sin \phi_\perp) \sinh \eta_\perp ] \\
&&  u^{\eta_\perp} = \frac{\gamma}{\tau_\perp} [
 (v_x\cos
\phi_\perp + v_y \sin \phi_\perp)  \cosh \eta_\perp -\sinh
\eta_\perp ] \\
&& u^{\phi_\perp}
= \frac{\gamma}{\tau_\perp \sinh \eta_\perp} (v_y \cos \phi_\perp - v_x \sin \phi_\perp)  \\
&& u^z=\gamma v_z
\end{eqnarray}

The metric tensor associated with the
$(\tau_\perp,\eta_\perp,\phi_\perp,z)$ coordinates is
\begin{eqnarray} \label{eqn_metric}
&& g_{mn}=Diag(1,-\tau_\perp^2,-\tau_\perp^2 \sinh^2\eta_\perp ,-1) \nonumber \\
&& g^{mn}=Diag(1,-\frac{1}{\tau_\perp^2} , -\frac{1}{\tau_\perp^2
\sinh^2\eta_\perp} ,-1)
\end{eqnarray}
With the metric above, it can be explicitly checked that $u^m u_m
= g_{mn} u^m u^n=1$.

For the covariant derivatives in the hydrodynamic equations
Eq.(\ref{eqn_hydro}), we will need the Affine connections
$\Gamma^{j}_{mn}= g^{jk}\Gamma_{kmn}=g^{jk} \frac{1}{2}(g_{km}\,
_{,\, n}+g_{kn}\, _{,\, m}-g_{mn}\, _{,\, k})$. In our case the
{\em non-vanishing} connections are:
\begin{eqnarray} \label{eqn_connection}
&& \Gamma^{\tau_\perp}_{\eta_\perp \eta_\perp}=\tau_\perp \, , \,
\Gamma^{\eta_\perp}_{\eta_\perp
\tau_\perp}=\Gamma^{\eta_\perp}_{\tau_\perp
\eta_\perp}=\frac{1}{\tau_\perp}
\nonumber \\
&& \Gamma^{\tau_\perp}_{\phi_\perp \phi_\perp}= \tau_\perp
\sinh^2\eta_\perp \, , \, \Gamma^{\phi_\perp}_{\tau_\perp
\phi_\perp}=\Gamma^{\phi_\perp}_{\phi_\perp
\tau_\perp}=\frac{1}{\tau_\perp} \nonumber \\
&& \Gamma^{\eta_\perp}_{\phi_\perp \phi_\perp}= -\sinh\eta_\perp
\cosh\eta_\perp \, , \, \Gamma^{\phi_\perp}_{\eta_\perp
\phi_\perp}=\Gamma^{\phi_\perp}_{\phi_\perp \eta_\perp} =
\frac{\cosh\eta_\perp}{\sinh\eta_\perp}
\end{eqnarray}

\section{2D Hubble Embedding and Derived Velocity Equation}

Now to find solutions in (1+3)D space-time, we first do an
explicit 2D Hubble embedding, which involves the following ansatz
for the transverse part of the velocity field:
\begin{eqnarray}
v_x=\frac{x}{t}=\tanh\eta_\perp \cos\phi_\perp \,\, , \,\,
v_y=\frac{y}{t}=\tanh\eta_\perp \sin\phi_\perp
\end{eqnarray}
which in the $(\tau_\perp,\eta_\perp,\phi_\perp,z)$ coordinates
translate nicely into:
\begin{eqnarray}
u^{\tau_\perp} =\frac{\gamma}{\cosh\eta_\perp} \,\, , \,\,
u^{\eta_\perp} = 0 \nonumber \\
u^{\phi_\perp} =0 \,\, , \,\, , u^z=\gamma v_z
\end{eqnarray}
The above can be further re-written as:
\begin{eqnarray} \label{eqn_2d_embedding}
 u^m = \bar \gamma (1,0,0,\bar v_z)
\end{eqnarray}
by introducing $\bar v_z=v_z\cosh\eta_\perp \, , \, \bar \gamma=
\sqrt{1-\bar v_z^2}$.

Now inserting the velocity field Eq.(\ref{eqn_2d_embedding})
together with the metric Eq.(\ref{eqn_metric}) and the E.o.S
Eq.(\ref{eqn_eos}) (which substitutes the $p$ by $\nu w$) into the
energy-momentum tensor in Eq.(\ref{eqn_tmunu}), we can explicitly
obtain all the non-vanishing components to be:
\be
&&T^{\tp\tp}=\left[-1+\frac{\gm^2}{\nu}\right]\, p \\
&&T^{\tp z}=\frac{\gm^2\vz}{\nu}\, p \\
&&T^{\eta\eta}=\frac{p}{\tau_\perp^2} \\
&&T^{\phi\phi}=\frac{p}{\tau_\perp^2\sinh^2\eta_\perp} \\
&&T^{zz}=\left[1+\frac{\gm^2\vz^2}{\nu}\right]\, p \ee

From the above, we can explicitly write down the hydrodynamic
equations below: \be && \left[-\nu\tp +\gm^2\tp
\right]\, p,_{\tp} +\gm^2\tp \vz\,  p,_{z} \nonumber
\\ && \quad +
\left[2\gm\tp(\gm,_{\tp}+\vz\gm,_{z})+\gm^2
(2+\tp\vz,_{z})\right]\, p=0 \\
&&p,_{\eta_\perp}=0\\
&&p,_{\phi_\perp}=0\\
&& \gm^2\vz\tp\,  p,_{\tp} +
\left[\gm^2\vz^2\tp+\nu \tp\right]\,  p,_{z}
 \nonumber \\
&&\quad
+\left[2\gm\vz(\gm,_{\tp}+\gm,_{z}\vz)\tau_\perp+\gm^2(\tp\vz,_{\tp}+2\vz(1+\tp\vz,_{z}))\right]
\, p =0 \ee The two equations about $\eta_\perp$ and $\phi_\perp$
are trivially solved by setting all fields independent of these
two variables, i.e. $p(x^m)=p(\tau_\perp,z)$ and
$\vz(x^m)=\vz(\tau_\perp,z)$. Then we can convert the remaining
two equations into the following: \be && (\ln p)_{,\tp}\equiv
{\mathcal F}[\tau_\perp,z]
\\
&&\quad =\frac{-\tp\vz(\vz{}_{,\tp}+\vz\vz{}_{,z})
+\nu\left[2(1-\vz^2)+2\tp\vz\vz{}_{,\tp}+\tp(1+\vz^2)\vz{}_{,z}\right]}
{(-1+\nu)\nu\tp(1-\vz^2)^2} \nonumber \\
&&(\ln p)_{,z} \equiv {\mathcal G}[\tau_\perp,z]
\\
&&\quad =\frac{\tp(\vz{}_{,\tp}+\vz\vz{}_{,z})
+\nu\left[-2\vz(1-\vz^2)-\tp(1+\vz^2)\vz{}_{,\tp}-2\tp\vz\vz{}_{,z}\right]}{(-1+\nu)\nu\tp(1-\vz^2)^2}
\nonumber \ee

The commutativity of partial derivative $(\ln
p)_{,\tau_\perp,z}=(\ln p)_{,z,\tau_\perp}$ leads to the following
constraint which is a derived equation only involving the velocity
field:

\be \label{constraint}
&&\tp^2[\vz{}_{,\tp,\tp}+2\vz{}_{,\tp}\vz{}_{,z}-2\vz^3(\vz{}_{,\tp,z}-\vz{}_{,z}^2)+2\vz(2\vz{}_{,\tp}^2+\vz{}_{,\tp,z}+\vz{}_{,z}^2)\nonumber\\
&&-\vz^4\vz{}_{,z,z}+\vz^2(-\vz{}_{,\tp,\tp}+6\vz{}_{,\tp}\vz{}_{,z}+\vz{}_{,z,z})]\nonumber\\
&&+\nu[2\vz^5-12\tp^2\vz^2\vz{}_{,\tp}\vz{}_{,z}-2\vz^3(2-2\tp\vz{}_{,z}+\tp^2(\vz{}_{,\tp}^2-2\vz{}_{,\tp,z}+\vz{}_{,z}^2))\nonumber\\
&&-2\vz(-1+2\tp\vz{}_{,z}+\tp^2(3\vz{}_{,\tp}^2+2\vz{}_{,\tp,z}+3\vz{}_{,z}^2))
+\tp\vz^4(2\vz{}_{,\tp}+\nonumber\\
&&\tp(\vz{}_{,\tp,\tp}+\vz{}_{,z,z}))-\tp(\vz{}_{,\tp}(2+4\tp\vz{}_{,z})+\tp(\vz{}_{,\tp,\tp}+\vz{}_{,z,z}))]\nonumber\\
&&=0
\ee


Once the above derived velocity equation has been solved, the
matter field can be obtained as:
\begin{eqnarray}
p =p_0 \cdot Exp\left [ \int_{{\tau_\perp}_0}^{\tau_\perp}
d{\tau_\perp}' {\mathcal F[{\tau_\perp}',z]} + \int_{z_0}^z dz'
{\mathcal G[{\tau_\perp},z']} \right ]
\end{eqnarray}
with $p_0$ being the value at arbitrary reference point
${\tau_\perp}_0 \, , \,  z_0$.

\section{Analytic Solutions with Transverse 2D Hubble Embedding and
Longitudinal Power Law Ansatz}

As an application of the derived velocity equation
(\ref{constraint}) in the previous section, we now find {\em all }
solutions with power law dependence on the variables $\tau_\perp$
and $z$. Within such ansatz of the velocity field \be
\label{eqn_z_ansatz} \vz=A \tau_\perp^B z^C \ee
Eq.(\ref{constraint}) can be expressed explicitly as:
 \be
&&(AC\nu-AC^2\nu) \, \tau_\perp^{2+B}z^{-2+C}\nonumber \\
&&+(-AB+AB^2+2A\nu-AB\nu-AB^2\nu) \, \tau_\perp^Bz^C \nonumber\\
&&+(A^3B+3A^3B^2-4A^3\nu-6A^3B^2\nu) \, \tau_\perp^{3B}z^{3C}\nonumber \\
&&+(2A^5\nu+A^5B\nu-A^5B^2\nu) \, \tau_\perp^{5B}z^{5C} \nonumber\\
&&+(4A^2BC-4A^2C\nu-8A^2BC\nu)\, \tau_\perp^{1+2B}z^{-1+2C}\nonumber \\
&&+(-A^3C+3A^3C^2-6A^3C^2\nu)\, \tau_\perp^{2+3B}z^{-2+3C}\nonumber \\
&&+(4A^4BC+4A^4C\nu-8A^4BC\nu)\, \tau_\perp^{1+4B}z^{-1+4C} \nonumber\\
&&+(A^5C+A^5C^2-A^5C\nu-A^5C^2\nu)\, \tau_\perp^{2+5B}z^{-2+5C} \,
=0 \ee

In the above equation, terms with various powers of $\tau_\perp,z$
(and only power terms) appear: to make all of them, either
mutually cancel (among terms with exactly the same $\tp,\rho$
powers) or vanish by respective coefficients, to eventually zero
is quite nontrivial. A thorough sorting of the sequences of
$\tp,\rho$ powers can exhaust all possibilities to satisfy this
algebraic equation. After enumerating all the possibilities, we
find only four sets of solutions: \be \label{eqn_powerlaw_results}
&&(i) \quad A=0,\; \text{with}\; x^2+y^2<t^2 \, ;\nonumber \\
&&(ii) \quad A=1, B=-1, C=1,\; \text{with} \; x^2+y^2+z^2<t^2 \, ; \nonumber \\
&&(iii) \quad A=1, B=1, C=-1,\; \text{with}\; x^2+y^2<t^2\; \text{and}\; x^2+y^2+z^2>t^2\, ;  \nonumber \\
&&(iv) \quad A=-3, B=-1, C=1,\; \text{with}\; \nu=\frac{1}{2}\; \text{and}\;  x^2+y^2+9z^2<t^2 \, . \nonumber \\
\ee The solution (i) is just the 2D Hubble expansion without
z-motion, the solution (ii) is actually 3D Hubble expansion, while
the remaining two are more nontrivial. The applicable kinematic
domain for each solution is given above in terms of $(t,x,y,z)$
coordinates. The solution (iii) composed of two causally
disconnected pieces: $z>\sqrt{t^2-x^2-y^2}$ and
$z<-\sqrt{t^2-x^2-y^2}$. It can be considered as another type of
explosion\footnote{Of course, the concepts of explosion and implosion
depend on the reference point. We will choose the origin as the reference
point if not further specified}, with the longitudinal flow being
``anti-Hubble''. The solution (iv) corresponds to the
special EOS $\nu=1/2$, which will be further discussed in
Section-6.

\section{Analytic Solutions with Transverse 2D Hubble Embedding and
Nontrivial Longitudinal Flow}

In this Section, we try to seek more interesting solutions with
nontrivial longitudinal flow, i.e. beyond the simple power law
ansatz. Instead of directly going to the second order differential
equation in Eq.(\ref{constraint}), we return to
Eqs.(\ref{p_eqns_1})(\ref{p_eqns_4}) noting the fact that both are
homogeneous equations. Motivated by this fact and by the form of
2D and 3D Hubble flows, we start with the following joint ansatz
for both the pressure and the velocity field:
\begin{eqnarray}
&&p(\tp,z)=\tp^{-\frac{a}{1-\nu}}g(\frac{z}{\tp}) \\
&&\vz(\tp,z)=f(\frac{z}{\tp})  \end{eqnarray} The above ansatz
particularly features nontrivial dependence on the scaling
variable $\xi=\frac{z}{\tp}$ and extra nonscaling time structure
in the pressure.  It is interesting to note that 2D Hubble flow
corresponds to the trivial solution $f=0$ at $a=2$ while the 3D
Hubble flow corresponds to $f=\xi$ at $a=3$, both of which are
special cases of the above ansatz.

By inserting the above ansatz to (\ref{p_eqns_1}) and
(\ref{p_eqns_4}), we obtain two first order ordinary differential
equations involving $g(\xi)\equiv g(\frac{z}{\tp})$ and
$f(\xi)\equiv f(\frac{z}{\tp})$. Solving for $f'(\xi)$ and
$g'(\xi)$ from the equations (with $'$ meaning $\frac{d}{d\xi}$),
we obtain:

\be \label{eq_f}
&&f'(\xi)=-\frac{\nu(1-f(\xi)^2)(2-a+af(\xi)^2-2\xi f(\xi))}{(\nu-1)(f(\xi)-\xi)^2+\nu(\xi f(\xi)-1)^2} \\
&&g'(\xi)=-\frac{g}{1-\nu} \times \nonumber
\\ \label{eq_g}
&& \frac{-2f(\xi)+af(\xi)+2\nu f(\xi)-2a\nu f(\xi)+2\xi-a\xi-2\nu \xi+a\nu
\xi+a\nu \xi f(\xi)^2}{(\nu-1)(f(\xi)-\xi)^2+\nu(\xi f(\xi)-1)^2} \nonumber \\
\ee

We find that the Eq.(\ref{eq_f}) involves $f(\xi)$ only and thus
decouples from $g(\xi)$. So one may first focus on solving
(\ref{eq_f}) and with $f(\xi)$ obtained one can easily find
$g(\xi)$ from integrating Eq.(\ref{eq_g}) . Again one may check
that the 2D Hubble flow with $f=0$ at $a=2$ and the 3D Hubble flow
with $f=\xi$ at $a=3$ both satisfy the above equations.
We observe Eq.(\ref{eq_f}) and Eq.(\ref{eq_g}) are invariant under simultaneous
transformations: $\xi\rightarrow-\xi,\,f\rightarrow-f,\,g\rightarrow g$.
This looks like a manifestation of the time inversion symmetry,
i.e. $\tp\rightarrow-\tp$, in ideal hydrodynamics,
where dissipation terms are absent in stress energy tensor. However,
we note $\tp>0$ by definition. The symmetry can be interpretted as
parity: $z\rightarrow-z$.
 It leads to
the conclusion that any solution to Eq.(\ref{eq_f}) and Eq.(\ref{eq_g}) is
accompanied by its parity inversion counterpart. We will identify such pairs
in explicit solutions later.

To solve (\ref{eq_f}), we can also consider it as (inversely)
determining $\xi=\xi(f)$ as a function of variable $f$. We can
then introduce a new function
\begin{eqnarray}
t(f)\equiv\frac{f \, \xi(f)-1}{f-\xi(f)}
\end{eqnarray}
and recast (\ref{eq_f}) into a differential equation for $t(f)$:
\begin{eqnarray}\label{eq_t}
&& \left[(1-f^2)(a\,f-2t+a\,t)\right]\times
\frac{dt}{df} \nonumber \\
&& \qquad- \left[(\ka^2-a)f+(\ka^2+2-a)t+(a-1)f\,
t^2+(a-3)t^3\right] =0
\end{eqnarray}
In the above we have replaced the E.o.S parameter $\nu$ by
$\kappa$ with the relation $\nu=\frac{1}{\ka^2+1}$($\ka^2\ge 1$
such that $\nu\le 1/2$). In what follows we try to seek solutions
$t(f)$ with two different types of ansatz.

\subsection{Solutions from Linear Polynomial Ansatz}

Motivated by the fact that all the coefficients in Eq.(\ref{eq_t})
are polynomials of both $t$ and $f$, we naturally start with {\em
polynomial ansatz} for the function $t(f)$. The simplest form is:
\begin{eqnarray}
t(f) = \alpha \, f + \beta
\end{eqnarray}
Inserting the ansatz into Eq.(\ref{eq_t}), we obtain
\begin{eqnarray}
&& C_0 + C_1 f + C_2 f^2 + C_3 f^3  = 0 \Rightarrow\\
&& \,\, 0=  C_0 = -\beta
[2-a+\kappa^2+(2-a)\alpha+(a-3)\beta^2] \nonumber \\
&& \,\, 0= C_1=  a-\kappa^2+(-2+2a-\kappa^2)\alpha+(a-2)\alpha^2
+(1-a)\beta^2 +(9-3a)\alpha \beta^2 \nonumber \\
&& \,\, 0= C_2 = - \beta \alpha [-4+3a+(3a-9)\alpha ] \nonumber \\
&& \,\, 0= C_3 = -\alpha (1+\alpha) [a+(a-3)\alpha] \nonumber
\end{eqnarray}
The resulting four algebraic equations can be easily solved, with
all solutions listed below:\\
\textbf{I.} $\alpha=0,\beta=0$ leading to
\begin{eqnarray} \label{eqn_solution_1}
t=0 \,\, \to \,\,  f= \frac{1}{\xi} \,\, , \,\, \text{with} \quad
a=\kappa^2
\end{eqnarray}
\textbf{II.} $\alpha=\kappa^2,\beta=0$ leading to
\begin{eqnarray} \label{eqn_solution_2}
t=\frac{a\, f}{3-a} \,\, \to \,\, f= \frac{(1+\kappa^2)\xi \pm
\sqrt{(1+\kappa^2)^2\xi^2-4\kappa^2}}{2\kappa^2} \,\, , \,\, \text{with}
\quad a=\frac{3\kappa^2}{1+\kappa^2}
\end{eqnarray}
The first solution coincides with the ``anti-Hubble'' solution
found in previous section. The second solution is a new class of
solutions. We will return to a detailed discussion of it at the
end of this section. There are also nonphysical solutions which we
discard, for example the solution $\alpha=-1,\beta=0$ to the set
of algebraic equations leads to $t=-f$ and thus $f=\pm 1$ which
does not comply with the physical constraint $|f|<1$.

One in principle can use various more general ansatz, for example:
(a) polynomials beyond the linear one we used; (b) polynomial
types with non-integer powers. These would typically be more
complicated but may lead to more solutions. One class we examined
is the form $t=\alpha f^\gamma+\beta$ with $\gamma$ being any
positive real number, which leads to no more solutions other than
the ones found above with $\gamma=1$.

\subsection{Solutions from Nonlinear Ansatz}

In this subsection we go beyond the linear polynomial ansatz and
show one example of nonlinear ansatz, which lead to interesting
new solutions. Motivated by the $1-f^2$ term in Eq.(\ref{eq_t}),
we are led to the following nonlinear ansatz:
\begin{eqnarray} \label{eqn_nonlinear_ansatz}
t(f) = \alpha \, f + \beta + \frac{\rho f + \lambda}{1-f^2}
\end{eqnarray}
Furthermore we consider only specific solutions with $a=3$ which
has the advantage of ``killing'' the $t^3$ term in the
Eq.(\ref{eq_t}). Inserting this ansatz into the equation for
$t(f)$ we obtain:
\begin{eqnarray}
&& \frac{1}{1-f^2} \, \left[ D_0 + D_1 f + D_2 f^2 + D_3 f^3 + D_4 f^4 + D_5 f^5 \right]  = 0 \Rightarrow\\
&& \,\, 0 = D_0 = (\beta+\lambda) (1-\kappa^2+\alpha+\rho) \nonumber \\
&& \,\, 0 = D_1 = 3-\kappa^2+(4-\kappa^2)\alpha+\alpha^2
-2\beta(\beta+\lambda)+(4-\kappa^2)\rho+ \rho^2+2\alpha \rho \nonumber \\
&& \,\, 0= D_2 = \beta (\kappa^2-1-6\alpha-3\rho)+3\lambda(2-\alpha) \nonumber \\
&& \,\, 0= D_3 = \kappa^2-3+(\kappa^2-7)\alpha-4\alpha^2
+2\beta^2+3\rho-4\alpha\rho \nonumber \\
&& \,\, 0= D_4 = 5 \alpha \beta \nonumber \\
&& \,\, 0= D_5 = 3\alpha (1+\alpha) \nonumber
\end{eqnarray}
By solving the above set of algebraic equations we found all the
nontrivial and physical solutions listed below:\\
\textbf{III.} $\alpha=0,\beta=2,\rho=-2,\lambda=-2$ leading to
\begin{eqnarray}\label{eqn_solution_3}
t=-\frac{2f}{1-f} \,\, \to \,\, f= \frac{1}{2-\xi} \,\, , \,\,
\text{with} \quad a=3 \,\, \text{and} \,\, \kappa=1
\end{eqnarray}
\textbf{IV.} $\alpha=0,\beta=-2,\rho=-2,\lambda=2$ leading to
\begin{eqnarray}\label{eqn_solution_4}
t=-\frac{2f}{1+f} \,\, \to \,\, f= -\frac{1}{2+\xi} \,\, , \,\,
\text{with} \quad a=3 \,\, \text{and} \,\, \kappa=1
\end{eqnarray}
\textbf{V.} $\alpha=0,\beta=1,\rho=-1,\lambda=-1$ leading to
\begin{eqnarray}\label{eqn_solution_5}
t=-\frac{f}{1-f} \,\, \to \,\, f= \frac{1\pm
\sqrt{5-4\xi}}{2(\xi-1)} \,\, , \,\, \text{with} \quad a=3 \,\, \text{and} \,\,
\kappa=2
\end{eqnarray}
\textbf{VI.} $\alpha=0,\beta=-1,\rho=-1,\lambda=1$ leading to
\begin{eqnarray}\label{eqn_solution_6}
t=-\frac{f}{1+f} \,\, \to \,\, f= \frac{1\pm
\sqrt{5+4\xi}}{2(\xi+1)} \,\, , \,\, \text{with} \quad a=3 \,\, \text{and} \,\,
\kappa=2
\end{eqnarray}
\textbf{VII.} $\alpha=0,\beta=0,\rho=-1,\lambda=0$ leading to
\begin{eqnarray}\label{eqn_solution_7}
&&t=-\frac{f}{1-f^2} \,\, \to \,\,
f=\left\{\begin{array}{l}
\frac{2}{3\xi}\left[
1-\zeta(\xi)-\bar\zeta(\xi)\right]\\
\frac{2}{3\xi}\left[
1+\frac{1+i\sqrt{3}}{2}\zeta(\xi)+\frac{1-i\sqrt{3}}{2}\bar\zeta(\xi)\right]\\
\frac{2}{3\xi}\left[
1+\frac{1-i\sqrt{3}}{2}\zeta(\xi)+\frac{1+i\sqrt{3}}{2}\bar\zeta(\xi)\right]
\end{array}
\right. \\
&&\text{with} \,\, a=3 \,\, \text{and} \,\, \kappa=\sqrt{6} \,\nonumber\\
&&\text{and} \,\,
\zeta(\xi)=\left[-1+\frac{27}{16}\xi^2+
i\sqrt{1-(1-\frac{27}{16}\xi^2)^2}\right]^{1/3} \nonumber\\
&&\quad \quad\, \bar\zeta(\xi)=\left[-1+\frac{27}{16}\xi^2-
i\sqrt{1-(1-\frac{27}{16}\xi^2)^2}\right]^{1/3}\nonumber
\end{eqnarray}

The case for general $a\ne 3$ with the same ansatz
(\ref{eqn_nonlinear_ansatz}) is also investigated, in which case a
set of 10 algebraic equations result. Many possibilities arise,
and we analyzed most of them and found no more solutions, but few
of the possibilities lead to really complicated situation that is
not analytically tractable. Though an exhaustive investigation of
general $a$ is not achieved, it seems most likely there is no
solutions other than those listed above with $a=3$. There are
certainly many more possible ansatz which could be tested in
future works.

\subsection{Discussion of Nontrivial Solutions}

In this section we choose to discuss properties and applicable kinematic domains
of solutions \textbf{II},\textbf{III},\textbf{IV},\textbf{V},\textbf{VI},\textbf{VII}.

We start with solution-\textbf{II}. It is subject to the physical constraint
 $|f|<1$ and reality condition of $f$.
The applicable kinematic domain of the solution-\textbf{II} given
in Eq.(\ref{eqn_solution_2}) is given below (where we already
replace $\xi$ by the original coordinate variables $z/\tp$ and
also replace $\kappa^2$ by $(1-\nu)/\nu$):

\be
&& \vz = f(\frac{z}{\tp}) \nonumber\\
&&p=\text{constant}\times\tp^{-3}\frac{f^{\frac{1-3\nu}{1-\nu}}}{(1-f^2)^{\frac{2-3\nu}{1-\nu}}} \nonumber\\
&&\text{domain 1:}\; f=\frac{z/\tp+\sqrt{z^2/\tp^2-4\nu(1-\nu)}}{2(1-\nu)}\nonumber\\
&&\text{with}\;2\sqrt{\nu(1-\nu)}<\frac{z}{\tp}< 1\\
&&\text{domain 2:}\; f=\frac{z/\tp+\sqrt{z^2/\tp^2-4\nu(1-\nu)}}{2(1-\nu)}\nonumber\\
&&\text{with}\;\frac{z}{\tp}< -1\\
&&\text{domain 3:}\; f=\frac{z/\tp-\sqrt{z^2/\tp^2-4\nu(1-\nu)}}{2(1-\nu)}\nonumber\\
&&\text{with}\;-1<\frac{z}{\tp}< -2\sqrt{\nu(1-\nu)}\\
&&\text{domain 4:}\; f=\frac{z/\tp-\sqrt{z^2/\tp^2-4\nu(1-\nu)}}{2(1-\nu)}\nonumber\\
&&\text{with}\;\frac{z}{\tp}> 1 \ee

In principle, any combination out of the four cases is an allowed
solution. Here we will treat them as four independent solutions as
there is no overlap in their applicable domains, i.e.they are
causally disconnected from each other. Note solutions in domain 1 and domain 3
are related by parity inversion. The same is true for
solutions in domain 2 and domain 4.
The transverse part of the
flow in all cases is always exploding 2D Hubble by embedding. The
longitudinal flow in $z$ direction, as indicated by the sign of
$f$, is also exploding outward in all four cases, since we have
$f>0$ for $\frac{z}{\tp}>0$ and $f<0$ for $\frac{z}{\tp}<0$. The
pressure drops as $\tp$ increases keeping $\frac{z}{\tp}$
constant. The solutions in some sense are like shock waves: at
each given time moment $\tp$, the flow is in certain domain of the
full space; while with changing $\tp$, the spatial shape of the
flow profile (in $z$) does not change but only  has its overall
size scale up with $\tp$.

Next we turn to solutions-\textbf{III} and \textbf{IV}. We first note
they are related by parity inversion. Therefore, we will concentrate on
solution-\textbf{III} only. With physical constraint $|f|<1$ applied,
solution-\textbf{III} looks like:

\be
&&\vz=f(\frac{z}{\tp})\nonumber\\
&&p=\text{constant}\times\tp^{-6}\frac{1+1/f}{(1/f-1)^3}\nonumber\\
&&\text{with}\;\nu=\frac{1}{2}\\
&&\text{domain 1:}\; f=\frac{1}{2-z/\tp}\nonumber\\
&&\text{with}\;\frac{z}{\tp}>3\\
&&\text{domain 2:}\; f=\frac{1}{2-z/\tp}\nonumber\\
&&\text{with}\;\frac{z}{\tp}<1
\ee

Similar to solution-\textbf{II}, solution-\textbf{III} also contains two
causally disconnected pieces. If we shift the scaling variable:
$z/\tp\to z/\tp-2$,
which amounts to choosing $z/\tp=2$ as the reference point.
The shifted solution $f=-\frac{\tp}{z}$ represents imploding fluid.

The other pair of solutions related by parity inversion is
\textbf{V} and \textbf{VI}. We will
focus on solution-\textbf{V}. In terms of original coordinate
variables, the solution with physical constraint applied is shown as follows:

\be
&&\vz=f(\frac{z}{\tp})\nonumber\\
&&p=\text{constant}\times\tp^{-15/4}\frac{1}{(1/f-1)^{15/8}(1+1/f)^{5/8}}\nonumber\\
&&\text{with}\;\nu=\frac{1}{5}\\
&&\text{domain 1:}\; f=\frac{1+\sqrt{5-4z/\tp}}{2(z/\tp-1)}\nonumber\\
&&\text{with}\;\frac{z}{\tp}< -1\\
&&\text{domain 2:}\; f=\frac{1-\sqrt{5-4z/\tp}}{2(z/\tp-1)}\nonumber\\
&&\text{with}\;\frac{z}{\tp}< 1
\ee

The transverse flow is again exploding by embedding, while the longitudinal part
differs from solution-\textbf{II}. Note $f$ does not change sign,
solution in domain 1 corresponds to fluid moving in the negative $z$ direction,
while solution in domain 2 corresponds to fluid moving in the positive $z$
direction. The pressure drops as $\tp$ increases keeping $\frac{z}{\tp}$
constant. They can be interpretted as shock waves with infinite extension.

Finally, we discuss solution-\textbf{VII}. Under the physical constraint
$|f|<1$ and reality condition, we are left with:

\be
&&\vz=f(\frac{z}{\tp})\nonumber\\
&&p=\text{constant}\times\tp^{-7/2}(1/f^2-1)^{-7/4} \nonumber\\
&&\text{with}\;\nu=\frac{1}{7} \nonumber\\
&&\text{domain 1:}\; f=f_1\equiv\frac{2\tp}{3z}\left[
1-\zeta(z/\tp)-\bar\zeta(z/\tp)\right]\nonumber\\
&&\text{with}\;\frac{z}{\tp}>0\;\text{or}\;\frac{z}{\tp}<0\\
&&\text{domain 2:}\; f=f_2\equiv\frac{2\tp}{3z}\left[
1+\frac{1+i\sqrt{3}}{2}\zeta(z/\tp)+\frac{1-i\sqrt{3}}{2}\bar\zeta(z/\tp)\right]\nonumber\\
&&\text{with}\;0<\frac{z}{\tp}<1\;\text{or}\;-1<\frac{z}{\tp}<0
\ee

Let us first comment on the reality of the solution: If
$1-(1-\frac{27}{16}\xi^2)^2>0$, i.e. the square root
in the definition of $\zeta$ and $\bar\zeta$ is real, we can
specify the argument of the square bracket such that $\zeta$ and $\bar\zeta$
are complex conjugate to each other, which guarantees the reality of $f_1$ and
$f_2$. If the square root becomes imaginary, only $f_1$ remains real.
The parity inversion simply maps $f_1$ and $f_2$ to themselves.
It is also worth noting that solution in domain 1 and domain 2 contain
discontinuity at $\frac{z}{\tp}=0$. They can be combined to give
two smooth solutions at $\frac{z}{\tp}=0$:

\be
\label{plusz}
&&f=\left\{\begin{array}{ll}
f_1& z/\tp<0\\
f_2& 0<z/\tp<1
\end{array}
\right. \\
\label{minusz}
&&f=\left\{\begin{array}{ll}
f_2& -1<z/\tp<0\\
f_1& z/\tp>0
\end{array}
\right.
\ee

This solution is similar to solution-\textbf{V}. It can also be interpretted
as shock wave with infinite extension.
(\ref{plusz}) corresponds to fluid moving in the positive $z$ direction,
while (\ref{minusz}) corresponds to fluid moving in the negative $z$
direction.

\section{General Scaling Solutions}

In this paper and in previous approaches, quite a few solutions
have been found with simple scaling forms for velocity field, it
is thus of interest to investigate all possible solutions with the
following simple scaling form velocity field (in $t,x,y,z$
coordinates):
\begin{equation}\label{scaling}
v_x = \alpha_x \frac{x}{t}\, , \, v_y = \alpha_y \frac{y}{t} \, ,
\, v_z = \alpha_z \frac{z}{t}
\end{equation}
The three constants $\alpha_{x,y,z}$ can in principle be different
corresponding to anisotropic scaling solutions. These forms
include many of the known solutions (cyclic in $x,y,z$):\\
1) $\alpha_x =1$, $\alpha_{y,z}=0$ --- Hwa-Bjorken or 1D Hubble
flow;\\
2) $\alpha_{x,y} =1$, $\alpha_{z}=0$ --- 2D Hubble flow;\\
3) $\alpha_{x,y,z} =1$ --- 3D Hubble flow;\\
4) setting $\alpha_z=1$ leads to a 1D Hubble embedding in Liao and
Koch paper \cite{Liao:2009zg}, in which
a scaling solution $\alpha_{x,y}=-1$ with $\nu=1/2$ has been found;\\
5) setting $\alpha_{x,y}=1$ leads to a 2D Hubble embedding in the
present paper, in which another scaling solution $\alpha_{z}=-3$
with $\nu=1/2$ has been found.\\
Beyond the above, it should be mentioned that generalization of
spherically symmetric scaling flow in arbitrary (1+d)D has been
done in \cite{Nagy:2007xn}.

It would be of great interest to exhaust all solutions with the
above simple scaling ansatz. To do that, we simply submit the flow
field into the hydrodynamics equation in (\ref{eqn_hydro}) (and
using simple flat coordinates $(t,\vec x)$), which leads to the
following equations for $p(t,x,y,z)$:
\begin{eqnarray}
\label{p_eqns_1} && \left[ h\,\left(t^2-\nu\, h \right)\right] \,
p_{,t} + \left[ h\, t x \alpha_x \right]\, p_{,x} + \left[ h\, t y
\alpha_y\right]\, p_{,y} + \left[ h\, t z \alpha_z\right]\, p_{,z}
\nonumber \\
&& \qquad +\, \left[ t \left(\alpha_s t^2-\zeta_x \alpha_x^2 x^2 -
\zeta_y \alpha_y^2 y^2
  - \zeta_z \alpha_z^2 z^2   \right) \right]\, p =0 \\
\label{p_eqns_2} &&  \left[ h t x \alpha_x \right]\, p_{,t} +
\left[h \left(\alpha_x^2 x^2+\nu h\right)
  \right]\,
  p_{,x} + \left[h \alpha_x \alpha_y x y \right]\, p_{,y} +\left[ h \alpha_x \alpha_z x z\right]\,
  p_{,z}\nonumber \\
&& \qquad + \, \left[\alpha_x x \left(\kappa_x t^2 - \lambda_x
\alpha_x^2 x^2
  - \xi^-_x \alpha_y^2 y^2 - \xi^+_x \alpha_z^2 z^2\right) \right] \, p
  =0 \\
\label{p_eqns_3} &&  \left[ h t y \alpha_y \right]\, p_{,t} +
\left[h \alpha_y \alpha_x y x  \right]\, p_{,x}
  + \left[h  \left(\alpha_y^2 y^2+\nu h\right)
  \right]\,
  p_{,y}  +\left[ h \alpha_y \alpha_z y z\right]\,
  p_{,z}\nonumber \\
&& \qquad + \, \left[\alpha_y y \left(\kappa_y t^2 - \xi^+_y
\alpha_x^2 x^2 - \lambda_y \alpha_y^2 y^2
   - \xi^-_y \alpha_z^2 z^2\right) \right] \, p
  =0 \\
\label{p_eqns_4}  &&    \left[ h t z \alpha_z \right]\, p_{,t} +
\left[h \alpha_z \alpha_x z x  \right]\, p_{,x}
 +\left[ h \alpha_z \alpha_y z y\right]\,
  p_{,y} + \left[h  \left(\alpha_z^2 z^2+\nu h\right)
  \right]\,
  p_{,z}  \nonumber \\
&& \qquad + \, \left[\alpha_z z \left(\kappa_z t^2 - \xi^-_z
\alpha_x^2 x^2  - \xi^+_z \alpha_y^2 y^2 - \lambda_z \alpha_z^2
z^2
   \right) \right] \, p
  =0
\end{eqnarray}
In the above, we have introduced a few constants (as certain
combinations of $\alpha_{x,y,z}$):
\begin{eqnarray}
&&\alpha_s=\alpha_x+\alpha_y+\alpha_z \\
&&\zeta_{x,y,z}=2+\alpha_s - 2 \alpha_{x,y,z}\\
&&\lambda_{x,y,z}=1+\alpha_s-\alpha_{x,y,z} \\
&&\kappa_{x,y,z} = -1 + \alpha_s+\alpha_{x,y,z} \\
&&\xi^{\pm}_{x,y,z} = 1+2\alpha_{x,y,z} \pm
(\alpha_{y,z,x}-\alpha_{z,x,y})
\end{eqnarray}
and also a coordinate-dependent function
\begin{eqnarray}
h(x^\mu)=t^2-\alpha_x x^2 - \alpha_y y^2 - \alpha_z z^2
\end{eqnarray}

The above derived equations (\ref{p_eqns_1},\ref{p_eqns_2},
\ref{p_eqns_3},\ref{p_eqns_4}) can be recast into the following
concise form:
\begin{eqnarray}
{\mathcal{M}}^{\mu \nu}\, \partial_\nu (\mathbf{ln p}) = B^\mu
\end{eqnarray}
from which we can explicitly solve out
\begin{eqnarray}
 \partial_\nu (\mathbf{ln p}) = \mathcal{M}^{\mathbf{-1}}_{\mu
 \nu} \,
B^\mu
\end{eqnarray}
The full expressions are somewhat long and not included here. In
analogy to Section.3, we can apply the compatibility condition,
i.e. the commutativity of partial derivatives, which give six
constraint equations on $\alpha_x$, $\alpha_y$ and $\alpha_z$.
Among these constraints, only four are independent. In practice,
we find it easier to solve the constraint equations by starting
from $(ln p)_{,x,y}=(ln p)_{,y,x}$ and then deal others
subsequently. After tedious calculations, it has been found that
an exhaustive list of all possible solutions include the following
(all solutions
valid subject to cyclic $x\to y\to z \to x$): \\
\textbf{I}. $\alpha_x=1$, $\alpha_{y,z}=0$ --- Hwa-Bjorken or 1D Hubble flow;\\
\textbf{II}. $\alpha_{x,y} =1$, $\alpha_{z}=0$ --- 2D Hubble flow;\\
\textbf{III}. $\alpha_{x,y,z} =1$ --- 3D Hubble flow;\\
\textbf{IV}. $\alpha_x=\alpha_y$ and
$1+\alpha_x+\alpha_y+\alpha_z=0$ with
$\nu=1/2$; \\
\textbf{V}. $\alpha_x=0$ and $1+\alpha_x+\alpha_y+\alpha_z=0$ with $\nu=1/2$. \\
The applicable kinematic region for all of the above solutions is
simply $\alpha_x^2 x^2+\alpha_y^2 y^2+\alpha_z^2 z^2<t^2$, which
comes from the constraint that the flow velocity shall be less
than the speed of light. We notice that there are {\em no more}
solution corresponding to general E.o.S parameter $\nu$ except the
well-known Hubble flows in various dimensions. While at $\nu=1/2$
(which somehow is a saturating extreme), we obtain two classes of
solutions, with the solution \textbf{IV} including the solutions
found previously as special cases and the solution \textbf{V} as a
completely new class. Note the condition
$1+\alpha_x+\alpha_y+\alpha_z=0$ implies the matter is
imploding($\alpha_i<0$) in at least one of the directions. The
Solution \textbf{IV} and \textbf{V} represent implosion, but could
contain one or two directions that are exploding($\alpha_i>0$).

\section{Summary}

In summary, we have investigated various analytic (1+3)D solutions
to relativistic ideal hydrodynamic equations based on known
low-dimensional scaling solutions. We first studied solutions with
transverse 2D Hubble Embedding, and derived a single ODE
Eq.(\ref{constraint}) for the remaining longitudinal velocity
field. Using this equation we found all solutions with transverse
2D Hubble embedding and power law ansatz (\ref{eqn_z_ansatz}) for
the remaining longitudinal velocity field, as listed in
Eqs.(\ref{eqn_powerlaw_results}). To find solutions with
transverse 2D Hubble embedding but nontrivial longitudinal flows
beyond power law ansatz, is both very interesting and extremely
hard, and nevertheless we managed to find a few such solutions
corresponding to different physical conditions, as listed in
Eqs.(\ref{eqn_solution_1},\ref{eqn_solution_2},\ref{eqn_solution_3},\ref{eqn_solution_4},
\ref{eqn_solution_5},\ref{eqn_solution_6},\ref{eqn_solution_7}).
The last question we studied is all possible scaling solutions
with the general scaling form in Eq.(\ref{scaling}) for the
velocity field, to which the complete answer was also found and
listed. These analytic solutions describe various examples of
explosion, implosion, and shock-wave like processes in
relativistic hydrodynamics. Though the initial motivation came
from successful numerical hydrodynamic description of heavy ion
collisions, the solutions presented in this paper may not directly
apply to those but provide useful general indications for similar
explosions in heavy ion collisions and also certain astrophysical
processes. Admittedly, a lot more work would be needed to reach a
better and closer analytic description of the hydrodynamic
phenomena relevant for today's relativistic heavy ion collisions.

\vspace{0.25in}

{\em \textbf{Acknowledgements:}} We are grateful to Volker Koch
and Edward Shuryak for discussions. The work of S.L. is supported
by by the US-DOE grants DE-FG02-88ER40388 and DE-FG03- 97ER4014.
The work of J.L. is supported by the Director, Office of Energy
Research, Office of High Energy and Nuclear Physics, Divisions of
Nuclear Physics, of the U.S. Department of Energy under Contract
No. DE-AC02-05CH11231.


\begin{thebibliography}{99}

\bibitem{Landau_book}
L.~D.~Landau and E.~M.~Lifshitz, {\it ``Fluid Mechanics''}, 2nd
ed., Butterworth-Heinemann, 1987.

\bibitem{Weinberg_book}
S.~Weinberg, {\it ``Gravitation and Cosmology: Principles and
Applications of the General Theory of Relativity''}, John Wiley \&
Sons, 1972.

\bibitem{Landau_Khalatnikov}
L.~D.~Landau, Izv. Akad. Nauk SSSR Ser. Fiz. {\bf 17}, 51(1953).
I.~M.~Khalatnikov, Zhur. Eksp. Teor. Fiz. {\bf 27}, 529(1954).

\bibitem{ES_book}
E.~Shuryak, {\it ``The QCD Vacuum, Hadrons and Superdense
Matter''}, 2nd ed., World Scientific Publishing Company, 2004.


\bibitem{Shuryak:1978ij}
  E.~V.~Shuryak,
  Phys.\ Lett.\  B {\bf 78}, 150 (1978)
  [Sov.\ J.\ Nucl.\ Phys.\  {\bf 28}, 408.1978\ YAFIA,28,796 (1978\ YAFIA,28,796-808.1978)].

\bibitem{Ollitrault:1992bk}
  J.~Y.~Ollitrault,
  Phys.\ Rev.\  D {\bf 46}, 229 (1992).

\bibitem{Teaney:2000cw}
  D.~Teaney, J.~Lauret and E.~V.~Shuryak,
  Phys.\ Rev.\ Lett.\  {\bf 86} (2001) 4783
  [arXiv:nucl-th/0011058].
\bibitem{Kolb:2000fha}
  P.~F.~Kolb, P.~Huovinen, U.~W.~Heinz and H.~Heiselberg,
  Phys.\ Lett.\  B {\bf 500} (2001) 232
  [arXiv:hep-ph/0012137].

\bibitem{Voloshin:2008dg}
  S.~A.~Voloshin, A.~M.~Poskanzer and R.~Snellings,
  [arXiv:0809.2949 [nucl-ex]].

\bibitem{Heinz:2009xj}
 P.~F.~Kolb and U.~W.~Heinz,
  [arXiv:nucl-th/0305084].
  U.~W.~Heinz,
  [arXiv:0901.4355 [nucl-th]].
   D.~A.~Teaney,
  [arXiv:0905.2433 [nucl-th]].
P.~Romatschke,
  [arXiv:0902.3663 [hep-ph]].


\bibitem{Schaefer:2009dj}
  T.~Schaefer and D.~Teaney,
  [arXiv:0904.3107 [hep-ph]].

\bibitem{Liao:2009gb}
  J.~Liao and V.~Koch,
  [arXiv:0909.3105 [hep-ph]].

\bibitem{Liao:2009ni}
  J.~Liao and V.~Koch,
  Phys.\ Rev.\ Lett.\  {\bf 103} (2009) 042302
  [arXiv:0902.2377 [nucl-th]].

\bibitem{Koch:2009wk}
  V.~Koch,
  [arXiv:0908.3176 [nucl-th]].


\bibitem{Shuryak_review}
E.~V.~Shuryak, Prog. Part. Nucl. Phys.{\bf 53}, 273 (2004);
  Prog.\ Part.\ Nucl.\ Phys.\  {\bf 62}, 48 (2009).
  [arXiv:0807.3033 [hep-ph]].

\bibitem{sQGP_review}
  M.~Gyulassy and L.~McLerran,
  Nucl.\ Phys.\  A {\bf 750} (2005) 30
  [arXiv:nucl-th/0405013].
  E.~V.~Shuryak,
  Nucl.\ Phys.\  A {\bf 750}, 64 (2005)
  [arXiv:hep-ph/0405066].

\bibitem{Liao:2006ry}
  J.~Liao and E.~Shuryak,
  Phys.\ Rev.\  C {\bf 75}, 054907 (2007);
  [arXiv:hep-ph/0611131].
    Phys.\ Rev.\ Lett.\  {\bf 101}, 162302 (2008);
  [arXiv:0804.0255[hep-ph]].
  J.~Liao and E.~V.~Shuryak,
  Phys.\ Rev.\  D {\bf 73} (2006) 014509
  [arXiv:hep-ph/0510110].
  J.~Liao and E.~V.~Shuryak,
  Nucl.\ Phys.\  A {\bf 775} (2006) 224
  [arXiv:hep-ph/0508035].
  J.~Liao and E.~Shuryak,
  Phys.\ Rev.\  C {\bf 77} (2008) 064905
  [arXiv:0706.4465 [hep-ph]].
  J.~Liao and E.~Shuryak,
  Phys.\ Rev.\ Lett.\  {\bf 102} (2009) 202302
  [arXiv:0810.4116 [nucl-th]].
  J.~Liao and E.~Shuryak,
  [arXiv:0804.4890 [hep-ph]].
  J.~Liao and E.~Shuryak,
  [arXiv:0809.2419 [hep-ph]].

\bibitem{Chernodub:2006gu}
  M.~N.~Chernodub and V.~I.~Zakharov,
  Phys.\ Rev.\ Lett.\  {\bf 98}, 082002 (2007).
  [arXiv:hep-ph/0611228].





\bibitem{Hwa}
R.~C.~Hwa, Phys.\ Rev.\ D {\bf 10}, 2260(1974).

\bibitem{Bjorken}
J.~D.~Bjorken, Phys.\ Rev.\ D{\bf 27}, 140(1983).




\bibitem{Nagy:2007xn}
  M.~I~Nagy, T.~Cs\"org\H{o} and M.~Csan\'ad,
  Phys.\ Rev.\  C {\bf 77}, 024908 (2008).
  [arXiv:0709.3677 [nucl-th]].



\bibitem{Bialas:2007iu}
  A.~Bialas, R.~A.~Janik and R.~B.~Peschanski,
  Phys.\ Rev.\  C {\bf 76}, 054901 (2007).
  [arXiv:0706.2108 [nucl-th]].

\bibitem{Pratt:2008jj}
  S.~Pratt,
  Phys.\ Rev.\  C {\bf 75}, 024907 (2007)
  [arXiv:nucl-th/0612010].

\bibitem{Wong:2008ex}
  C.~Y.~Wong,
  Phys.\ Rev.\  C {\bf 78}, 054902 (2008);
  [arXiv:0808.1294 [hep-ph]].
  [arXiv:0809.0517 [nucl-th]].

\bibitem{Csorgo:2002bi}
  T.~Cs\"org\H{o}, F.~Grassi, Y.~Hama and T.~Kodama,
  Heavy Ion Phys.\  A {\bf 21}, 63 (2004)
  [Acta Phys.\ Hung.\  A {\bf 21}, 63 (2004)].
  [arXiv:hep-ph/0204300].
\bibitem{Sinyukov:2004am}
  Yu.~M.~Sinyukov and I.~A.~Karpenko,
  Acta Phys.\ Hung.\  A {\bf 25}, 141 (2006).
  [arXiv:nucl-th/0506002].

\bibitem{Biro:1999eh}
  T.~S.~Bir\'o,
  Phys.\ Lett.\  B {\bf 474}, 21 (2000)
  [arXiv:nucl-th/9911004];
  Phys.\ Lett.\  B {\bf 487}, 133 (2000)
  [arXiv:nucl-th/0003027].

\bibitem{Peschanski:2009tg}
  R.~Peschanski and E.~N.~Saridakis,
  [arXiv:0906.0941 [nucl-th]].

\bibitem{Liao:2009zg}
  J.~Liao and V.~Koch,
  Phys.\ Rev.\  C {\bf 80}, 034904 (2009)
  [arXiv:0905.3406 [nucl-th]].




\end{thebibliography}
\end{document}